\documentclass[twocolumn,showpacs]{revtex4} 
\usepackage{amsmath, amssymb, graphicx} 
 
\newcommand{\ud}{\mathrm{d}} 
\newcommand{\ic}{\mathit{i}} 
\newcommand{\re}{\mathrm{ Re }} 
\newcommand{\Ai}{\mathrm{ Ai}} 
\newcommand{\sgn}{\mathrm{sgn}} 
 
\begin{document} 
 
\title{On the origin of the exponential decay of the Loschmidt echo in 
  integrable systems} 
 
\author{R\'{e}my Dubertrand$^{1,2}$ and Arseni Goussev$^{3,4}$} 
 
\affiliation{$^1$Universit\'{e} de Toulouse; UPS, Laboratoire de 
  Physique Th\'{e}orique (IRSAMC); F-31062 Toulouse, 
  France\\ $^2$CNRS; LPT (IRSAMC); F-31062 Toulouse, France 
  \\ $^3$Department of Mathematics and Information Sciences, 
  Northumbria University, Newcastle Upon Tyne, NE1 8ST, United 
  Kingdom\\ $^4$Max Planck Institute for the Physics of Complex 
  Systems, N{\"o}thnitzer Stra{\ss}e 38, D-01187 Dresden, Germany} 
 
\date{\today} 
 
\begin{abstract} 
  We address the time decay of the Loschmidt echo, measuring
  sensitivity of quantum dynamics to small Hamiltonian perturbations,
  in one-dimensional integrable systems. Using semiclassical analysis,
  we show that the Loschmidt echo may exhibit a well-pronounced regime
  of exponential decay, alike the one typically observed in quantum
  systems whose dynamics is chaotic in the classical limit. We derive
  an explicit formula for the exponential decay rate in terms of the
  spectral properties of the unperturbed and perturbed Hamilton
  operators and the initial state. In particular, we show that the
  decay rate, unlike in the case of the chaotic dynamics, is directly
  proportional to the strength of the Hamiltonian
  perturbation. Finally, we compare our analytical predictions against
  the results of a numerical computation of the Loschmidt echo for a
  quantum particle moving inside a one-dimensional box with
  Dirichlet-Robin boundary conditions, and find the two in good
  agreement.
\end{abstract} 
 
 
\maketitle 
 
\section{Introduction} 
 
Even a tiny perturbation of the Hamiltonian of a quantum systems may 
significantly alter its time evolution. Changes in the dynamics of the 
corresponding unperturbed and perturbed systems can be conveniently 
quantified in terms of the Loschmidt echo (LE) that is defined as 
\begin{equation} 
  M(t) = \big| m(t) \big|^2 \,, 
\label{LE-definition} 
\end{equation} 
with 
\begin{equation} 
  \label{def_m} 
  m(t)=\langle \Psi_{\lambda_2} (t) | \Psi_{\lambda_1} (t) \rangle\ 
\end{equation} 
being the overlap between two quantum states, $| \Psi_{\lambda_1} (t) 
\rangle$ and $| \Psi_{\lambda_2} (t) \rangle$, resulting from the same 
initial state $| \Psi_{\lambda_1} (0) \rangle = | \Psi_{\lambda_2} (0) 
\rangle = | \Psi_0 \rangle$ in the course of the time evolution under 
different Hamilton operators, $H_{\lambda_1}$ and $H_{\lambda_2}$ 
respectively. The LE equals 1 at $t=0$ and is typically smaller than 1 
at $t > 0$. Over the past two decades, the time decay of the LE has 
been addressed, both experimentally and theoretically, in a large 
variety of quantum systems with non-trivial, complex dynamics, and has 
been proven an invaluable tool in understanding dynamical properties 
of these systems. A vast body of literature on the subject is 
summarized in review articles 
\cite{GPSZ06Dynamics,JP09Decoherence,GJPW12Loschmidt}. 
 
Most of the attention on the LE has been directed at quantum systems 
whose dynamics is chaotic in the classical limit. This has to do with 
the existence of a parametric regime, commonly referred to as the 
Lyapunov regime, in which the LE of a chaotic quantum system decays 
exponentially in time, with the decay rate given by the average 
Lyapunov exponent of the underlying classical system 
\cite{JP01Environment,JSB01Golden,CPW02Decoherence,CPJ04Universality}. The 
Lyapunov regimes makes the LE a valuable tool for identifying 
signatures of chaotic behaviour in the dynamics of quantum systems. 
 
In integrable (regular) systems, unlike in chaotic systems, the decay
of the LE remains far less understood. Two robust decay regimes have
been established analytically and observed in numerical
simulations. The first regime is characterized by a Gaussian decay of
the LE, $M(t) \sim \exp(-\mathrm{constant} \times t^2)$. It occurs
under sufficiently weak Hamiltonian perturbations and takes place on a
time scale inversely proportional to the perturbation strength
\cite{Pro02General}. The second regime exhibits an algebraic decay of
the LE, $M(t) \sim t^{-3 d / 2}$, with $d$ being the dimensionality of
the system. The algebraic decay occurs when the Hamiltonian
perturbation is sufficiently strong and varies rapidly along a typical
classical trajectory of the system \cite{JAB03Anomalous}. In general,
however, the time decay of the LE in integrable systems is
non-monotonic, may be accompanied by revivals \cite{SL03Recurrence}
and intervals of temporary freeze \cite{PZ03Quantum}, and may exhibit
sharp minima and maxima on short time scales \cite{Goussev11}.

Surprisingly, numerical simulations have shown that, in addition to 
the Gaussian and algebraic decays, integrable systems may also exhibit 
exponential decay of the LE \cite{WH05Quantum,WCL07Stability}. This 
finding may seem unexpected as the exponential decay of the LE is 
normally regarded as a hallmark of chaotic dynamics. To our knowledge, 
there is currently no theoretical model able to quantitatively explain 
exponential decay of the LE in integrable systems. 
 
In this paper, we show analytically that exponential decay of the LE 
may occur even in the simplest, one-dimensional integrable 
systems. For such systems, we derive an explicit formula giving the 
exponential decay rate in terms of spectral characteristics of the 
unperturbed and perturbed Hamilton operators and the initial state. In 
particular, we show that the decay rate is directly proportional to 
the strength of the Hamiltonian perturbation. This linear dependence 
is to be compared with the corresponding rate-vs-strength dependence 
in the case of chaotic dynamics: Under weak perturbations, the decay 
rate is quadratic in the perturbation strength (the Fermi-golden-rule 
regime) \cite{JSB01Golden}, while under strong perturbations, the 
decay rate is independent of the perturbation strength (the Lyapunov 
regime) \cite{JP01Environment}. Finally, using an example system -- a 
quantum particle moving inside a one-dimensional box with 
Dirichlet-Robin boundary conditions -- we compare our analytical 
prediction against the exact (numerically computed) LE, and find the 
two in good agreement.

\section{Theory of the exponential decay in one-dimensional systems} 
 
Let $H_{\lambda}$ be a family of integrable one-dimensional 
Hamiltonians, parametrised by some parameter $\lambda$. The 
eigenvalues and eigenvectors of $H_{\lambda}$ are respectively denoted 
by $E_{\lambda}(n)$ and $\left|\psi_{\lambda}(n)\right>$, where $n$ 
stands for the quantum number of the one-dimensional system. 
 
In what follows, we consider a parametric regime of asymptotically 
small perturbations, in which $\lambda_1 = \lambda$ and 
$\lambda_2=\lambda+\epsilon$ with $|\epsilon| \ll |\lambda|$. The LE 
amplitude $m(t)$, defined by Eq.~(\ref{def_m}), can be expressed as a 
double sum over the eigenstates of both ${H}_{\lambda}$ and 
${H}_{\lambda + \epsilon}$. Using the diagonal approximation, 
corresponding to the leading order perturbation theory, one arrives at 
the standard, single sum approximation to the LE amplitude 
\cite{Peres84}: 
\begin{eqnarray} 
 \label{diagm} m(t)&\simeq& \sum_{n} 
    \big| \left< \psi_{\lambda}(n) | \Psi_0 \right> \big|^2\ e^{ \ic
      [E_{\lambda+\epsilon}(n)-E_{\lambda}(n)]t/\hbar}\ . 
\end{eqnarray} 
Infinite sums, mathematically similar to the one in Eq.~(\ref{diagm}), 
are commonly encountered in studies of time-domain autocorrelation 
functions of quantum wave packets \cite{Nau90,Rob04}. 
 
We analyze the sum in Eq.~(\ref{diagm}) in the semiclassical limit of 
high energies. Motivated by the study of quantum revivals of the 
autocorrelation function in Ref.~\cite{Nau90}, we consider an initial 
state that is a linear superposition of a large number of highly 
excited states: 
\begin{equation} 
  \label{initstate} 
  \big| \left< \psi_{\lambda}(n) | \Psi_0 \right> \big|^2 \simeq 
  \frac{1}{\sqrt{2\pi(\Delta n)^2}} 
  \exp\left[\frac{-(n-n_0)^2}{2(\Delta n)^2}\right] 
\end{equation} 
with 
\begin{equation} 
  \label{init2} 
  1 \ll \Delta n \ll n_0\ . 
\end{equation} 
We note that, in view of the condition (\ref{init2}), the initial 
state is normalized to unity, $\sum_n \big| \langle \psi_{\lambda} (n) 
| \Psi_0 \rangle \big|^2 \simeq 1$. 
 
We now assume that, for large enough values of the quantum number, the 
energy levels can be asymptotically expanded as 
\begin{equation} 
  E_{\lambda}(n) \simeq h(n) + g(\lambda) + f(\lambda) n^\nu \,,  
\quad n\gg 1\ .\label{Escl}
\end{equation} 
Here, the terms $h(n)$ and $g(\lambda)$ are respectively functions of $n$ and 
$\lambda$ only. The term $f(\lambda) n^{\nu}$ depends on both $n$ and 
$\lambda$, and is the first, leading-order term of the expansion (in 
powers of $1/n$) whose coefficient depends on $\lambda$.  The exponent 
$\nu$ is a nonzero real number. We note that it is the term 
$f(\lambda) n^{\nu}$ that will play a crucial role in the following 
analysis. 

Expansion~(\ref{Escl}) approximates the spectrum of a rather broad
class of quantum systems, whose energy levels in the semiclassical
(high-energy) regime can be expanded into a power series in the
quantum number. Some examples are the motion of a quantum particle
confined to the potential well $V(x)=V_0 |x|^\alpha$ with any real
positive $\alpha$, including the harmonic oscillator, and the radial
motion of an electron in the hydrogen-like atom. Furthermore, we only
consider smooth perturbations that do not change the power series
structure of the energy spectrum.

In view of Eq.~(\ref{Escl}), the
phase difference in Eq.~(\ref{diagm}) becomes
\begin{equation} 
\label{diffEn} 
  E_{\lambda+\epsilon}(n)-E_{\lambda}(n) \simeq \epsilon g'(\lambda) + 
  \epsilon f'(\lambda) n^\nu \,, 
\end{equation} 
where the prime denotes the derivative. The time scale for which this 
approximation holds coincides with the range of validity of the 
perturbative regime. Then, expanding the right-hand side of 
Eq.~(\ref{diffEn}) around $n_0$, we obtain 
\begin{align} 
   E_{\lambda+\epsilon}&(n)-E_{\lambda}(n) \simeq \epsilon g'(\lambda) 
   \nonumber\\ &+\epsilon f'(\lambda) n_0^\nu \, \sum_{k=0}^{\infty} 
   \frac{\Gamma(\nu+1)}{\Gamma(\nu-k+1) k!} \left( \frac{n-n_0}{n_0} 
   \right)^k \,, 
\label{diffEn2} 
\end{align} 
where $\Gamma$ denotes the Euler Gamma function generalising the factorial. 
 
It is instructive to note that some well known results for the LE 
decay can be recovered by truncating the infinite sum in 
Eq.~(\ref{diffEn2}) and substituting the truncated sum into 
Eq.~(\ref{diagm}). The Gaussian decay regime is obtained if one only 
keeps terms up to the linear one in $(n-n_0)/n_0$, i.e., the two terms 
corresponding to $k=0$ and $k=1$. Retaining additionally the quadratic 
term, $k=2$, leads to an algebraic modification of the variance in the 
Gaussian regime. As we will show below, the cubic term, $k=3$, gives 
rise to a new asymptotic decay regime, which follows the 
(algebraically modified) Gaussian decay, and in which the LE decays 
essentially exponentially in time. 
 
Substituting Eq.~(\ref{initstate}) and Eq.~(\ref{diffEn2}), truncated 
at $k=3$, into Eq.~(\ref{diagm}), and replacing the sum by the 
corresponding integral, we obtain 
\begin{equation} 
  \label{m3} 
  m(t) \simeq e^{i \phi} \int_{-\infty}^{\infty} \exp\left( \ic a n^3 
  -b n^2 +\ic c n\right) \frac{\ud n}{ \sqrt{2\pi (\Delta n)^2}} 
\end{equation} 
with $\phi = \epsilon g'(\lambda)t / \hbar$, 
\begin{eqnarray} 
  a&=&\frac{\nu(\nu-1)(\nu-2)}{6} n_0^{\nu-3}\tau\nonumber \,, \\ 
  b&=&\dfrac{1}{2(\Delta n)^2} - \frac{\ic}{2} \nu(\nu-1) n_0^{\nu-2}\tau \nonumber \,, \\ 
  c&=& \nu n_0^{\nu-1} \tau \,, 
\label{listm3} 
\end{eqnarray} 
and 
\begin{equation} 
  \tau = \frac{\epsilon f'(\lambda) t}{\hbar} \,. 
\label{tau} 
\end{equation} 
The integral in Eq.~(\ref{m3}) can be evaluated using the identity 
\cite{LAS96} 
\begin{align} 
\int_{-\infty}^{\infty} e^{\ic a x^3-bx^2 +\ic c x}\ud x 
=&\frac{2\pi}{|3a|^{\frac{1}{3}}} 
\exp\left(\frac{bc}{3a}+\frac{2b^3}{27 a^2}\right) \nonumber\\ &\times 
\Ai\left[\frac{\sgn(a)}{|3a|^{\frac{1}{3}}}\left(\frac{b^2}{3a}+c\right) 
  \right] \label{airy} \ , 
\end{align} 
valid for real $a$ and $c$, and $\re\ b>0$. Here, $\Ai$ denotes the
Airy function and $\sgn(x)$ is the sign function, defined as
$\sgn(x)=1$ for $x>0$ and $\sgn(x)=-1$ for $x<0$. Focusing on time
scale beyond the validity range of the Gaussian decay regime,
\begin{equation} 
  \tau \gg  n_0^{2-\nu} (\Delta n)^{-2} \,, 
\label{large_tau} 
\end{equation} 
and using the identity (\ref{airy}) in Eq.~(\ref{m3}), we obtain, to 
the leading order in $n_0^{2-\nu} (\Delta n)^{-2} / \tau$, 
\begin{align} 
  |m(t)| \simeq \, &\sqrt{\dfrac{2\pi}{(\Delta n)^2}} \left|
  \dfrac{2}{\nu(\nu-1)(\nu-2) n_0^{\nu-3} \tau} \right|^{\frac{1}{3}}
  \nonumber\\ &\times \exp\left[ -\frac{1}{(\nu-1)(\nu-2)^2} \left(
    \frac{n_0}{\Delta n} \right)^2 \right] \nonumber\\ &\times
  \Ai\left[ \sgn \left( \frac{\nu-3}{\nu-1} \right)
    \frac{|\nu-3|}{|\nu-1|^{\frac{1}{3}}} \left( \frac{|\nu| n_0^\nu
      \tau}{2 (\nu-2)^2} \right)^{\frac{2}{3}} \right]\,.
\label{mod_m} 
\end{align} 
We now assume that $\nu < 1$, which guaranties the argument of the
Airy function in Eq.~(\ref{mod_m}) be strictly positive.  Then, in
view of conditions (\ref{init2}) and (\ref{large_tau}), we use the
large argument asymptotics of the Airy function \cite{gradshteyn},
\begin{equation} 
  \label{asymptAiry} 
  \Ai(x) \simeq \dfrac{1}{2\sqrt{\pi} x^{\frac{1}{4}}} \exp \left( 
  -\frac{2}{3} x^{\frac{3}{2}} \right) \,, \quad x \gg 1 \,, 
\end{equation} 
and obtain the following approximate expression for the LE: 
\begin{equation} 
  M(t) \simeq \frac{A}{t} e^{-\gamma t} \,, 
\label{exp_decay} 
\end{equation} 
with 
\begin{equation} 
  \gamma = \frac{2 |\nu|}{3(\nu - 2)^2} \sqrt{ \frac{|\nu - 3|^3}{|\nu 
      - 1|} } \, \frac{|\epsilon f'(\lambda)| n_0^\nu}{\hbar} 
\label{decay_rate} 
\end{equation} 
and 
\begin{align} 
  A = &\frac{n_0^{2-\nu} (\Delta n)^{-2} \hbar}{\sqrt{\nu^2 (\nu-1) 
      (\nu-3)} \, |\epsilon f'(\lambda)|} \nonumber\\ &\times 
  \exp\left[ -\frac{2}{(\nu-1)(\nu-2)^2} \left( \frac{n_0}{\Delta n} 
    \right)^2 \right] \,. 
\label{prefactor} 
\end{align} 
 
Equations~(\ref{exp_decay}--\ref{prefactor}) constitute the main 
analytical result of the present paper. Equation~(\ref{exp_decay}) 
shows that the LE may decay exponentially in time (not taking into 
account an algebraic prefactor) even in quantum systems whose 
counterpart classical dynamics is not chaotic, and as simple as that 
of a one-dimensional conservative system. 
 
\section{Particle in a box with Dirichlet-Robin boundary conditions} 
 
\begin{figure}[ht] 
\includegraphics[width=3in]{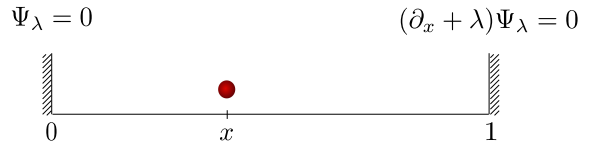} 
\caption{(Color online) Particle in a one-dimensional box with Dirichlet-Robin 
  boundary conditions.} 
\label{fig1} 
\end{figure} 
 
We now illustrate our theory by considering dynamics of a quantum 
particle trapped inside a one-dimensional box, $0 < x < 1$. The 
particle state $| \Psi(t) \rangle$ evolves according to the 
free-particle Schr\"{o}dinger equation, 
\begin{equation} 
  \left( \ic \partial_t + \partial_x^2 \right) \langle x |
  \Psi_{\lambda}(t) \rangle = 0 \,,
\label{Robin-01} 
\end{equation} 
subject to Dirichlet boundary condition at $x=0$ and Robin boundary 
condition at $x=1$ (see Fig.~\ref{fig1}), 
\begin{align} 
  &\langle x | \Psi_{\lambda}(t) \rangle \big|_{x=0} = 0 
  \,, \label{Robin-02a} \\ &(\partial_x + \lambda) \langle x | 
  \Psi_{\lambda}(t) \rangle \big|_{x=1} = 0 \,. \label{Robin-02b} 
\end{align} 
Hereinafter, we set $\hbar = 1$ and $m = 1/2$. We address the time 
decay of the LE, defined by Eq.~(\ref{LE-definition}), due to a small 
perturbation of the real-valued control parameter $\lambda$, i.e., 
$\lambda_1 = \lambda$ and $\lambda_2 = \lambda + \epsilon$ specify the 
unperturbed and perturbed systems respectively. We take the initial 
state, $| \Psi_{\lambda_1}(0) \rangle = | \Psi_{\lambda_2}(0) \rangle 
= | \Psi_0 \rangle$, to be given by a Gaussian wave packet 
\begin{equation} 
  \langle x | \Psi_0 \rangle = \left( \frac{1}{\pi \sigma^2} 
  \right)^{\frac{1}{4}} \exp \left( \ic p_0 (x - x_0) - \frac{(x - 
    x_0)^2}{2 \sigma^2} \right) \,. 
\label{Robin-02.5} 
\end{equation} 
Here, $x_0$ and $p_0$ correspond to the initial average position and
momentum of the particle respectively, and $\sigma$ quantifies the
spatial dispersion of the wave packet. Provided $0 < x_0 < 1$ and
$\sigma \ll x_0,(1-x_0)$, the entire probability density is well
localized within the box and the boundary conditions (\ref{Robin-02a})
and (\ref{Robin-02b}) are satisfied with exponentially high
accuracy. We use the values $x_0 = 0.5$ and $\sigma = 0.01$ in all our
numerical examples.
 
Eigenstates $| \psi_{\lambda}(n) \rangle$ and energy levels 
$E_{\lambda}(n) \equiv z_{\lambda}^2(n)$ of the system are determined 
by the equation 
\begin{equation} 
  \left[ \partial_x^2 + z_{\lambda}^2(n) \right] \langle x | 
  \psi_{\lambda}(n) \rangle = 0 \quad \mathrm{for} \quad n \in 
  \mathbb{N} \,, 
\label{Robin-03} 
\end{equation} 
along with the boundary conditions (\ref{Robin-02a}) and 
(\ref{Robin-02b}). Written out explicitly, the orthonormal eigenstates 
are 
\begin{equation} 
  \langle x | \psi_{\lambda}(n) \rangle = \sqrt{2} \left[ 1-\frac{\sin 
      \big(2z_{\lambda}(n)\big)}{2z_{\lambda}(n)} 
    \right]^{-\frac{1}{2}} \sin \big(z_{\lambda}(n) x\big) \,, 
\label{Robin-04} 
\end{equation} 
where the energy levels are determined by solutions of the 
transcendental equation 
\begin{equation} 
  z_{\lambda} (n) \cos z_{\lambda} (n) + \lambda \sin z_{\lambda} (n) 
  = 0 \,. 
\label{Robin-05} 
\end{equation} 
Equation~(\ref{Robin-05}) can be straightforwardly solved numerically, 
to a high degree of accuracy, for any desired range of $n$. For our 
purposes, it proved sufficient to compute the first 350 eigenstates 
and eigenlevels of the unperturbed and perturbed systems. 
 
In terms of the eigenstates and eigenlevels, the expression for the LE 
takes the form 
\begin{align} 
  M(t) = \Bigg| \sum_{n,m} &\langle \Psi_0 | \psi_{\lambda_2} (m) 
  \rangle \langle \psi_{\lambda_2} (m) | \psi_{\lambda_1} (n) \rangle 
  \nonumber \\ &\times \langle \psi_{\lambda_1} (n) | \Psi_0 \rangle 
  \, e^{ -\ic \big[ z_{\lambda_1}^2(n) - z_{\lambda_2}^2(m) \big] t} 
  \Bigg|^2 \,. \label{Robin-06} 
\end{align} 
Here, the amplitude of the overlap between the unperturbed and 
perturbed eigenstates can be written, using Eq.~(\ref{Robin-04}) 
together with the boundary conditions (\ref{Robin-02a}) and 
(\ref{Robin-02b}), as 
\begin{equation} 
  \langle \psi_{\lambda_2} (m) | \psi_{\lambda_1} (n) \rangle = 
  (\lambda_1-\lambda_2) \frac{\langle 1 | \psi_{\lambda_1} (n) \rangle 
    \langle \psi_{\lambda_2} (m) | 1 \rangle} {z_{\lambda_1}^2 (n) - 
    z_{\lambda_2}^2 (m)} \,. 
\label{Robin-07} 
\end{equation} 
We note that, for a sufficiently weak perturbation $\epsilon$, the 
right-hand side of Eq.~(\ref{Robin-07}) can be well approximated by 
the Kronecker's symbol $\delta_{m,n}$, and, consequently, the 
expression for the LE amplitude can be reduced to 
Eq.~(\ref{diagm}). More accurately, however, the value of the LE, 
$M(t)$, can be obtained by substituting Eqs.~(\ref{Robin-02.5}), 
(\ref{Robin-04}), and (\ref{Robin-07}) into Eq.~(\ref{Robin-06}), and 
numerically evaluating the overlap amplitudes $\langle \psi_{\lambda} 
(n) | \Psi_0 \rangle$ and the sums in Eq.~(\ref{Robin-06}). 
 
\begin{figure}[ht] 
\includegraphics[width=3.2in]{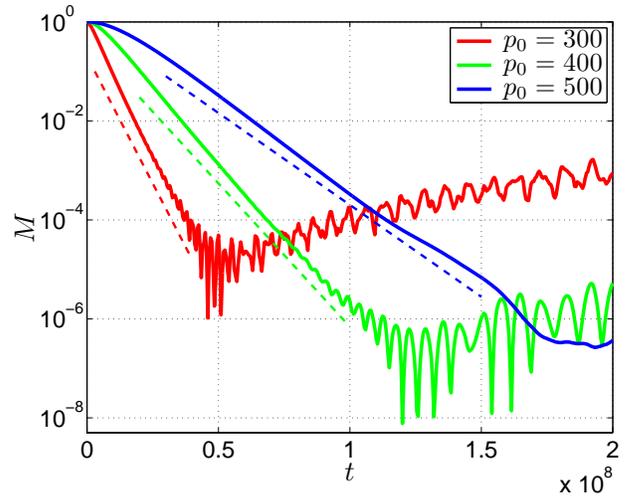} 
\caption{(Color online) Time decay of the LE for a particle moving inside a 
  one-dimensional box with Dirichlet-Robin boundary conditions. The 
  system, perturbation, and the initial state are characterized by the 
  following set of parameters: $\lambda = 1$, $\epsilon = 0.01$, $x_0 
  = 0.5$, $\sigma = 0.01$, and $p_0 = 300$ (red), 400 (green), 500 
  (blue). Solid curves represent the numerically exact LE. Straight 
  dashed lines of the corresponding color show the trend of the 
  exponential decay, $e^{-\gamma t}$, with $\gamma$ computed in 
  accordance with Eqs.~(\ref{Robin-15}). See text for details.} 
\label{fig2} 
\end{figure} 
 
Figure~\ref{fig2} presents three LE decay curves in the system with 
$\lambda = 1$ under the perturbation $\epsilon = 0.01$. The initial 
Gaussian state is specified by the average position $x_0 = 0.5$ and 
spatial dispersion $\sigma = 0.01$. Three values of the average 
momentum are considered: $p_0 = 300$ (red solid curve), $p_0 = 400$ 
(green solid curve), and $p_0 = 500$ (blue solid curve). All three 
curves exhibit well pronounced regions of nearly exponential decay 
featuring a LE drop over 3-4 orders of magnitude. It is interesting to 
note that the exponential decay persists over very long time 
intervals, corresponding to approximately $10^9 - 10^{10}$ periods of 
the underlying classical oscillation. 
 
In order to compare the numerically obtained LE decay curves with the 
analytical predictions of the previous section, we evaluate the 
exponential decay rate $\gamma$ in accordance with 
Eq.~(\ref{decay_rate}). To this end, we expand the energy eigenlevels, 
$E_{\lambda}(n) = z_{\lambda}^2 (n)$, into a series of the form 
(\ref{Escl}) for large $n$. We first write $z_{\lambda}$ in terms of a 
new quantity $\zeta_{\lambda}$ as 
\begin{equation} 
  z_{\lambda}(n) = \pi \left( n - \frac{1}{2} \right) + 
  \zeta_{\lambda}(n) \,. 
\label{Robin-08} 
\end{equation} 
Then, substituting Eq.~(\ref{Robin-08}) into Eq.~(\ref{Robin-05}), we obtain 
\begin{equation} 
  \left( \pi - n^{-1} \frac{\pi}{2} + n^{-1} \zeta_{\lambda} \right) 
  \sin \zeta_{\lambda} = n^{-1} \lambda \cos \zeta_{\lambda} \,. 
\label{Robin-09} 
\end{equation} 
Since $\zeta_{\lambda} \rightarrow 0$ as $n^{-1} \rightarrow 0$, we 
look for $\zeta_{\lambda}$ in the form of the following power series 
in $n^{-1}$: 
\begin{equation} 
  \zeta_{\lambda} = n^{-1} A_{\lambda} + n^{-2} B_{\lambda} + n^{-3} 
  C_{\lambda} + \mathcal{O}(n^{-4}) \,. 
\label{Robin-10} 
\end{equation} 
Substituting Eq.~(\ref{Robin-10}) into Eq.~(\ref{Robin-09}), expanding 
both sides of the resulting equation into power series in $n^{-1}$, 
and then matching the expansion coefficients on both sides of the 
equation, we find 
\begin{align} 
  A_{\lambda} &= \frac{\lambda}{\pi} \,, \label{Robin-11a} 
  \\ B_{\lambda} &= \frac{\lambda}{2 \pi} \,, \label{Robin-11b} 
  \\ C_{\lambda} &= \frac{\lambda}{4 \pi} - \frac{\lambda^2}{\pi^3} - 
  \frac{\lambda^3}{3 \pi^3} \,. \label{Robin-11c} 
\end{align} 
Then, using Eqs.~(\ref{Robin-10}--\ref{Robin-11c}) into 
Eq.~(\ref{Robin-08}), we get 
\begin{align} 
  z_{\lambda} (n) = &\pi n - \frac{\pi}{2} + \frac{\lambda}{\pi n} + 
  \frac{\lambda}{2 \pi n^2} \nonumber \\ &+ \left( \frac{\lambda}{4 
    \pi} - \frac{\lambda^2}{\pi^3} - \frac{\lambda^3}{3 \pi^3} \right) 
  n^{-3} + \mathcal{O} (n^{-4}) \,. 
\label{Robin-12} 
\end{align} 
Raising $z_{\lambda}$ to the second power, we obtain 
\begin{equation} 
  E_{\lambda} (n) = \pi^2 \left( n - \frac{1}{2} \right)^2 + 2 \lambda 
  - \left( 1 + \frac{2}{3} \lambda \right) \frac{\lambda^2}{\pi^2 n^2} 
  + \mathcal{O} (n^{-3}) \,. 
\label{Robin-13} 
\end{equation} 
Comparing Eq.~(\ref{Robin-13}) with Eq.~(\ref{Escl}), we conclude
that, for the system of a particle moving inside a one-dimensional
Dirichlet-Robin box,
\begin{equation} 
  \nu = -2 \quad \mathrm{and} \quad f(\lambda) = - \left( 1 + 
  \frac{2}{3} \lambda \right) \frac{\lambda^2}{\pi^2} \,. 
\label{Robin-14} 
\end{equation} 
Finally, a substitution of Eq.~(\ref{Robin-14}) into 
Eq.~(\ref{decay_rate}) leads to the following explicit formula for the 
decay rate: 
\begin{equation} 
  \gamma = \frac{5}{6 \pi^2} \sqrt{\frac{5}{3}} \, \frac{|(1+\lambda) 
    \lambda \epsilon|}{n_0^2} \,. 
\label{Robin-15} 
\end{equation} 
 
The value of $n_0$ can be computed numerically as $n_0 = \sum_n n 
\big| \langle \psi_{\lambda} (n) | \Psi_0 \rangle \big|^2$. For the 
three cases presented in Fig.~\ref{fig2}, we obtain $n_0 \simeq 96$ 
for the initial state with $p_0 = 300$ (red curve), $n_0 \simeq 128$ 
for $p_0 = 400$ (green curve), and $n_0 \simeq 160$ for $p_0 = 500$ 
(blue curve). A substitution of these values of $n_0$, along with 
$\lambda = 1$ and $\epsilon = 0.01$, yields the corresponding values 
of the decay rate $\gamma$. Dashed lines in Fig.~\ref{fig2} present 
the trend of the exponential decay, $e^{-\gamma t}$, for each of the 
three cases. It is evident that Eq.~(\ref{Robin-15}) (or, 
equivalently, Eq.~(\ref{decay_rate})) provides a very good estimate 
for the rate of the exponential decay of the LE. 
 
Finally, we note that the quantum number dispersion, $\Delta n$,
evaluated as $(\Delta n)^2 = \sum_n (n-n_0)^2 \big| \langle
\psi_{\lambda} (n) | \Psi_0 \rangle \big|^2$, approximately equals 23
for all three initial states considered in this section. This value is
consistent with condition (\ref{init2}), which is one of the central
assumptions of our theory.

\section{Discussion and conclusion} 

In this paper we have developed an analytical theory of the long-time
exponential decay of the Loschmidt echo (or fidelity) in
one-dimensional integrable quantum systems under the action of
integrable perturbations. In particular, we have shown that, in the
integrable case, the rate of the exponential decay $\gamma$ is
proportional to the perturbation strength $|\epsilon|$. This is to be
contrasted with the $\gamma$ versus $\epsilon$ dependence typically
observed in quantum systems whose dynamics is chaotic in the classical
limit: $\gamma \propto |\epsilon|^2$ for weak perturbation
(Fermi-golden-rule regime) and $\gamma \propto |\epsilon|^0$ for
strong perturbations (Lyapunov regime). It is interesting to note that
the linear dependence of the decay rate on the perturbation strength,
$\gamma \propto |\epsilon|^1$, has been also observed in a quantum
system with chaotic classical counterpart under the action of a local
perturbation \cite{wisniacki}. We think it is important to explore the
connection between this work and our results.

As a test of our analytical formulas we have applied our theory to the
system of a quantum particle moving inside a one-dimensional box with
Dirichlet boundary condition on one end and Robin boundary condition
on the other. The Robin control parameter was used to induce a
perturbation. We have shown the system to exhibit a well pronounced
regime of an exponential decay of the Loschmidt echo, and found a good
agreement between the analytically predicted and numerically observed
values of the decay rate.

Exponential decay of the Loschmidt echo in integrable systems has been
previously addressed by numerical simulations. The exponential decay
regime was clearly observed in Ref.~\cite{WH05Quantum}, however no
clear relation between the decay rate and the perturbation strength
was found. We attribute this to their particular choice of the initial
state, at variance with the semiclassical condition (\ref{init2}). In
Ref.~\cite{WCL07Stability}, the exponential decay regime was found to
be a transient from the short-time Gaussian to the long-time power law
decay. This is not in contradiction with our result. Our theory, in
addition to explaining the physical origin of the exponential decay
regime, provides an explicit formula for the decay rate, and, in
particular, gives the dependence of the decay rate on the perturbation
strength.

More generally, the current study emphasises the benefit of using the
Loschmidt echo as a tool for analyzing the long time asymptotics of
quantum dynamics in the semiclassical regime.  We have found that the
Loschmidt echo may decay exponentially even in one-dimensional quantum
systems. This can be related to the structural instability of the
corresponding classical integrable systems. A natural way to extend
our approach is to address higher dimensional systems. One of the
central challenges of such a study is the necessity to account for
quasi-degeneracies of the energy spectrum, generic in many dimensional
integrable systems.

\acknowledgments

R.D. acknowledges financial support by Programme Investissements
d'Avenir under the program ANR-11-IDEX-0002-02, reference
ANR-10-LABX-0037-NEXT (ENCOQUAM project). A.G. thanks EPSRC for
support under grant EP/K024116/1.

\end{document}